\documentclass[a4]{aa}
\usepackage{graphicx}
\usepackage{times}

\def \ref {\noindent\hangindent=1.0in\hangafter=1}

\def\ltsima{$\; \buildrel < \over \sim \;$}
\def\simlt{\lower.5ex\hbox{\ltsima}} 
\def\gtsima{$\; \buildrel > \over \sim \;$}
\def\simgt{\lower.5ex\hbox{\gtsima}} 
\def \integral{\emph{INTEGRAL}}
\def \3c454{3C~454.3}
\def \cgs{erg s$^{-1}$ cm$^{-2}$}

\begin{document}

\title{\emph{INTEGRAL} observations of the blazar 3C~454.3 in 
outburst\thanks{Based on observations obtained with \emph{INTEGRAL}, an 
ESA mission with instruments and science data center funded by ESA member 
states (especially the PI countries: Denmark, France, Germany, Italy, 
Switzerland, Spain, Czech Republic and Poland), and with the participation 
of Russia and the USA.}}

\author{E.~Pian\inst{1}, L.~Foschini\inst{2}, 
V.~Beckmann\inst{3,4},
S.~Soldi\inst{5,6},
M.~T\"urler\inst{5,6},
N.~Gehrels\inst{3},
G.~Ghisellini\inst{7},
P.~Giommi\inst{8},
L.~Maraschi\inst{9},
T.~Pursimo\inst{10},
C.M.~Raiteri\inst{11},
G.~Tagliaferri\inst{7}, 
M.~Tornikoski\inst{12},
G.~Tosti\inst{13},
A.~Treves\inst{14},
M.~Villata\inst{11},
P.~Barr\inst{15}\thanks{Deceased 19 October 2005.},
T.~J.-L. Courvoisier\inst{5,6},
G.~Di Cocco\inst{2},
R.~Hudec\inst{16},
L.~Fuhrmann\inst{11,13},
G.~Malaguti\inst{2},
M.~Persic\inst{1},
F. Tavecchio\inst{7},
R.~Walter\inst{5,6}
}

\institute{INAF, Osservatorio Astronomico di Trieste, Via G.B. Tiepolo 11, 
I-34131 Trieste, Italy
\and
INAF, IASF-Bologna, Via Gobetti 101, I-40129 Bologna, Italy
\and
NASA Goddard Space Flight Center, Code 661, Greenbelt, MD 20771, USA
\and
Joint Center for Astrophysics, Department of Physics, University of Maryland, 
Baltimore County, MD 21250, USA
\and
\integral\ Science Data Center, Chemin d'\'Ecogia 16, 1290 Versoix, Switzerland
\and
Observatoire de Gen\`eve, 51 Ch. des Maillettes, 1290 Sauverny, Switzerland 
\and
INAF, Osservatorio Astronomico di Brera, Via E. Bianchi, 46, I-23807 Merate 
(LC), Italy
\and
ASI Science Data Center, Via Galileo Galilei, I-00044 Frascati, Italy
\and
INAF, Osservatorio Astronomico di Brera, Via Brera 28, I-20121 Milano, Italy
\and
Nordic Optical Telescope, Apartado 474, E-38700 Santa Cruz de La Palma, Spain
\and
INAF, Osservatorio Astronomico di Torino, Via Osservatorio 20, I-10025 Pino 
Torinese (TO), Italy
\and
Mets\"ahovi Radio Observatory, Mets\"ahovintie 114, FIN-02540 Kylm\"al\"a, 
Finland
\and
Dipartimento di Fisica, University of Perugia, Via A. Pascoli, I-06123 Perugia,
Italy
\and
Dipartimento di Fisica e Matematica, University of Insubria, Via Valleggio 11, 
I-22100 Como, Italy
\and
ESA-ESTEC, RSSD, Keplerlaan 1, Postbus 299, 2200 AG Noordwijk, The Netherlands
\and
Astronomical Institute, Academy of Sciences of the Czech Republic,
CZ-251 65 Ondrejov, Czech Republic}

\authorrunning{E. Pian et al.} 
\titlerunning{3C~454.3 with \emph{INTEGRAL}} 

\offprints{E. Pian: \texttt{pian@oats.inaf.it}}   

\abstract{In Spring 2005, the blazar \3c454\ underwent a dramatic outburst 
at all wavelengths from mm to X-rays.  This prompted \integral\ 
observations, accomplished in 15-18 May 2005. The source was detected by 
the \integral\ instruments from 3 to 200 keV in a bright state ($\sim 5 
\times 10^{-10}$ \cgs), at least a factor of 2-3 higher than previously observed. 
This is one of the brightest blazar detections 
achieved by \integral.  During the 2.5 days of \integral\ monitoring, we 
detected a $\sim$20\% decrease in the hard X-rays (20-40 keV), indicating 
that we have sampled the decaying part of the flare.  The decrease is less 
apparent in the soft X-rays (5-15 keV). The simultaneous optical 
variations are weakly correlated with those at soft X-rays, and not 
clearly correlated with those at hard X-rays.  The spectral energy 
distribution exhibits two components, as typically seen in blazars, which 
can be modeled with synchrotron radiation and inverse Compton scattering 
occurring in a region external to the broad line region. 
}

\maketitle 

\section{Introduction}

Blazars exhibit the largest luminosities and multiwavelength variability 
amplitudes among active galactic nuclei (AGN), due to the highly 
relativistic regimes present in their jets, nearly aligned with the 
observer line of sight, and to ensuing strong aberration effects 
(\cite{cmu1995}, \cite {mhu1997}). Their broad-band spectra are typically 
non-thermal, and are produced most probably by synchrotron radiation at 
lower energies, and by inverse Compton (IC) scattering off synchrotron or 
external photons, at higher energies. The extreme conditions present in 
blazar jets make them bright and ideal targets for observations at high 
energies, from X-rays to the GeV range. The \integral\ satellite has so 
far provided an important contribution in this field by performing hard 
X-ray spectroscopy of a number of blazars, up to redshift $\sim$2 (3C~273, 
Courvoisier et al.  2003b; 3C~279, Collmar et al.  2004; PKS B1830-211, De 
Rosa et al. 2005; PKS~0716+714, S5~0836+710, Pian et al. 2005a; Beckmann 
et al. 2006).

The Flat Spectrum Radio-Quasar (FSRQ) \3c454\ ($z = 0.859$)  is a known, 
variable blazar, exhibiting strong emission lines in UV (e.g., Pian, 
Falomo, \& Treves 2005b), remarkable optical variability, radio 
superluminal motion and a radio and X-ray jet (Lobanov et al. 2000; 
Marshall et al. 2005), and keV (Worrall \& Wilkes 1990; Comastri et al. 
1997; Tavecchio et al. 2002), MeV (McNaron-Brown et al. 1995; Zhang, 
Collmar, \& Sch\"onfelder 2005)  and GeV (Mukherjee et al. 1997; Hartman 
et al. 1999)  radiation.  Its spectral energy distribution (SED) has a 
synchrotron peak at FIR frequencies and an IC peak at MeV 
energies (Ghisellini et al. 1998). In April 2005, the source was reported 
to be in an exceptionally high state in all bands from mm to X-rays 
(Tornikoski, priv. comm.; Fuhrmann et al.  2006, and references therein; 
Remillard 2005).  Fig.~1 shows the {\it RXTE} ASM light curve of \3c454\ 
from beginning of April to end of June 2005.  Between the end of April and 
May 10 the blazar flared, becoming the brightest AGN in the X-ray sky. 
This triggered the activation of observations with {\it Swift} (24 April, 
11, 17, 19 May, Giommi et al.  2006), {\it RXTE} (12 May), {\it Chandra} 
(19-20 May), and \integral\ (15-18 May).  Here are presented the results 
of the \integral\ observation, preliminarily reported in Foschini et al. 
(2005).

\section{Observations and data analysis}
\label{}

A Target-of-Opportunity observation of \3c454\ 
with \integral\ (\cite{cw2003}) 
started on May 15, 2005, at 18:40 and ended on May 18, 2005, at 09:46 UT. 
The on source time for IBIS/ISGRI (\cite{pu2003}; \cite{lll03}) and SPI 
(\cite{gv2003}) detectors was $207$ ks, and for IBIS/PICsIT 
(\cite{pu2003}; \cite{gdc03}) was $180$ ks. Using a searching radius of 3 
degrees in the JEM-X (\cite{nl2003}) data, the source is in the JEM-X 
field-of-view for a total of 59~ks.

The screening, reduction, and analysis of the \integral\ data have been 
performed using the \integral\ Offline Scientific Analysis (OSA) V.~5.1, 
publicly available through the \integral\ Science Data 
Center\footnote{http://isdc.unige.ch/index.cgi?Soft+download} (ISDC, 
\cite{tjlc2003a}).  The algorithms implemented in the software are 
described in Goldwurm et al.  (2003)  for IBIS, Westergaard et al. (2003) 
for JEM-X, and Diehl et al. (2003) for SPI. The IBIS/ISGRI, SPI and JEM-X 
data have been accumulated into final images.  For the spectral analysis 
we used the most recent matrices available in OSA V. 5.0 and V. 5.1. The 
OMC (\cite{jmm2003}) data, which have been acquired with a standard V-band 
Johnson filter, have been extracted with default settings, using a 3x3 
pixels binning, which is appropriate for point-like sources.  Since this 
corresponds to a box of $50^{\prime\prime}$ size, the measurement of the 
V-band flux is contaminated by two stars, both located to the North-East, 
at $\sim$14$^{\prime\prime}$ distance from the blazar. One has a brightness 
comparable to that of the blazar, the other is a factor of $\sim$13 
fainter.  Therefore, after having checked that both stars are not 
variable, we have corrected the blazar flux for their contributions, 
equivalent to a total magnitude of $V = 13.59 \pm 0.02$ (see Raiteri et 
al. 1998).

Two days before the start of the \integral\ campaign and during four days 
following its end we observed \3c454\ with the 2.5m Nordic Optical 
Telescope (NOT) at the Canary Islands once almost every night with VRI 
filters (Fig.~1), and, during the maximum of the optical emission, 
occurred around May 19.5 UT, also with UBJHK filters.  The reduction of 
these data followed standard procedures.  The blazar is also a target of 
extensive mm monitoring at the Mets\"ahovi radio research station.

\section{Results}
\label{}

The blazar was detected with all \integral\ instruments, except 
IBIS/PICsIT, in the coadded images of the full observation. By adopting a 
conservative estimate for the IBIS/ISGRI background, significant signal 
($\sim$10 mCrab) is detected up to $\sim$200 keV.  With the SPI instrument 
we detect \3c454\ up to the same energy. The source is not seen in 
individual pointings by JEM-X, therefore its position was fixed to allow 
spectral extraction. The detection is highly significant, at the level of 
$\sim$4 mCrab.  The IBIS/PICsIT 3-$\sigma$ upper limit is $1.3 \times 
10^{-10}$ \cgs\ in the 252-336 keV band (corresponding to $\sim$53 mCrab). 
During this observation, five other blazars were located in the IBIS field 
of view, but went undetected. Their 3$\sigma$ upper limits in the 20-40 
keV band are 1.5 mCrab (PKS~2250+1926), 2 mCrab (CTA~102), 3 mCrab 
(PKS~2209+236), 7 mCrab (PKS~2356+196), and 14 mCrab (3C~445).

The IBIS/ISGRI (20-200 keV), SPI (20-200 keV) and JEM-X (3-15 keV) spectra 
have been combined and analyzed with 
XSPEC V. 11.  A fit with a single power-law modified by Galactic 
absorption (fixed to N$_{HI} = 6.5 \times 10^{20}$ cm$^{-2}$, Dickey \& 
Lockman 1990, and consistent with that derived from the LAB survey, 
Kalberla et al. 2005) yields a photon index $\Gamma = 1.8 \pm 0.1$ ($N(E) 
\propto E^{-\Gamma}$) with a $\chi^2 = 90.6$ for 137 degrees of freedom. 
The constants of intercalibration between IBIS/ISGRI and the JEM-X and SPI 
instruments are $0.4 \pm 0.1$ and $1.0 \pm 0.2$, respectively, consistent 
with the expected values (Cadolle Bel et al. 2006; Lubi\'nski, Dubath, \& 
Paltani 2005). The observed flux in the 3-200 keV band is $5.45 \times 
10^{-10}$ \cgs.  The flux extrapolated to the band 252-336 keV is $6.1 
\times 10^{-11}$ \cgs, consistent with the IBIS/PICsIT upper limit 
reported above.

We have subdivided the observation in four intervals of $\sim$50 ks each 
and have evaluated the count rates in the 5-15 keV and 20-40 keV ranges 
from the JEM-X and IBIS/ISGRI data, respectively. The light curves and hardness ratios 
between the hard and soft X-ray fluxes are 
shown in Fig.~2.  A decrease of $\sim$20\% (considering the uncertainties) 
of the 20-40 keV flux is observed during the monitoring.  The 5-15 keV 
flux is less variable, although a flux drop of at least 10-15\% is 
observed in the third temporal bin. The hardness ratios 
suggest a marginally significant softening 
with time.  We have also attempted power-law fits of the combined JEM-X 
and IBIS/ISGRI spectra in the individual temporal bins.  Only the first 
two spectra have sufficient photon statistics for a meaningful fit.  The 
photon indices are $\Gamma = 1.65 \pm 0.25$ and $\Gamma = 
1.87^{+0.28}_{-0.26}$, confirming the softening suggestion, but both formally 
consistent with the average photon index.

The OMC corrected flux is consistent with the simultaneous photometry 
taken from ground-based telescopes (e.g., Fuhrmann et al. 2006; Villata et 
al. 2006), including our own NOT measurements. The OMC light curve is 
shown in Fig.~2.  Significant variations of up to $\sim$0.15 magnitudes in 
time scales of less than an hour are detected.  The correlation between 
the optical and X-ray variations is difficult to assess, because of the 
limited number and large uncertainty of the high energy data points, and 
different temporal coverage. When compared to the high energy light curves 
on a time scale similar to their resolution, the OMC light curve 
appears not correlated with the hard X-ray emission and somewhat
correlated with the soft X-rays.


\begin{figure}
\centering
\includegraphics[angle=0,scale=0.45]{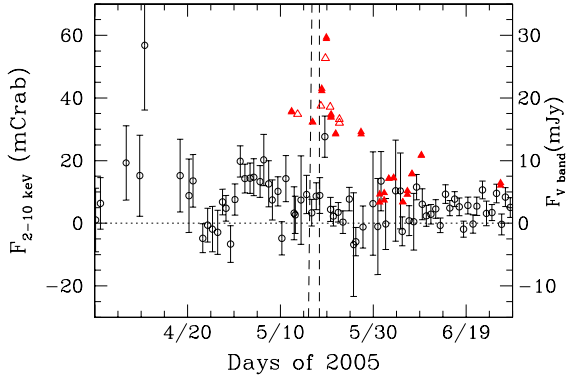}
\caption{{\it RXTE} ASM light curve of \3c454 in Spring 2005 (open 
circles) and optical V-band light curves from REM (filled triangles, 
Fuhrmann et al. 2006) and NOT (open triangles).  The dashed vertical lines 
mark the duration of the \integral\ observation}
\end{figure}


\begin{figure}
\centering
\includegraphics[angle=0,scale=0.6]{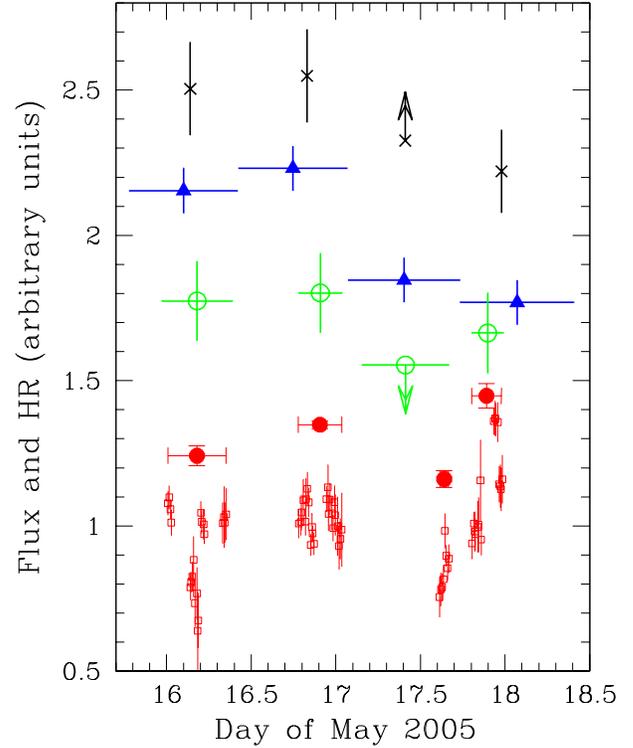}
\caption{\integral\ IBIS/ISGRI (20-40 keV, triangles), JEM-X (5-15 keV, 
open circles), and OMC V-band (with original time resolution, squares, and 
rebinned to match IBIS time resolution, filled circles)  light curves 
normalized to their respective averages of 1.3 counts~s$^{-1}$, 0.36 
counts~s$^{-1}$, and $\sim$20 mJy ($\langle V \rangle \simeq 13.5$ \rm \, 
mag).  Background-subtraction systematics have been taken into account in 
the IBIS error estimate.  The ratios between the IBIS and JEM-X fluxes
have been reported as crosses (normalized to their average of 3.87).  
For clarity, the binned OMC, JEM-X, IBIS light curves 
and the 
hardness ratios are shifted up by the additive constants 0.3, 0.7, 1 and 1.4, 
respectively}
\end{figure}


\begin{figure}
\centering
\includegraphics[angle=0,scale=0.4]{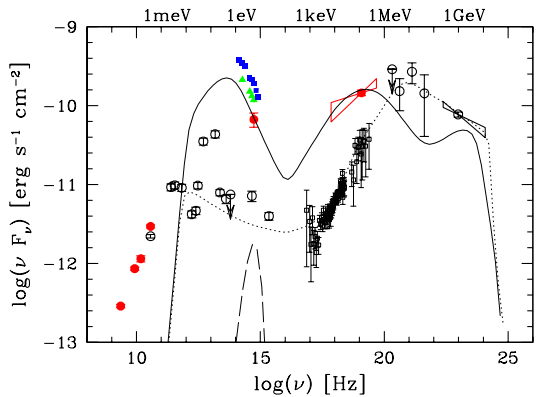}
\caption{Multiwavelength energy distribution of \3c454\ in observer
frame.  The filled symbols refer to observations in Spring 2005. The 
filled circles are data taken at 2.3 and 8.45 GHz (11-12 April 2005, 
Trushkin et al. 2005), 15 GHz (21 April 2005, {\tiny 
http://www.physics.purdue.edu/astro/MOJAVE/sourcepages/2251+158.shtml}), 
37 GHz (Mets\"ahovi, May-June 2005), optical wavelengths (our \integral\ 
OMC data of May 16.18), and X-rays (our \integral\ JEM-X, IBIS and SPI 
observations).  The filled triangles are optical/NIR data taken with REM 
simultaneously with the \integral\ campaign (May 16.9, Fuhrmann et al. 
2006).  The filled squares are our NOT data at optical/NIR wavelengths 
acquired on May 19.7. The open symbols represent observations taken at 
various epochs prior to the multiwavelength outburst of Spring 2005. The 
open circles are data taken at 37 GHz (Mets\"ahovi, early 2005), mm/FIR 
(Haas et al. 2004), optical (Fuhrmann et al. 2006, and references 
therein), UV (Pian et al. 2005b)  wavelengths, and at hard X- and 
gamma-rays (average COMPTEL, 0.75-30 MeV, and EGRET, 0.1-5 GeV, 
fluxes in the years 1991-1994, Zhang 
et al. 2005; Hartman et al. 1999). We also report the 1-$\sigma$ 
confidence limits of the \integral\ and EGRET spectra.  The open squares 
represent the X-ray spectrum measured by {\it BeppoSAX} in 2000 (Tavecchio 
et al. 2002). The NIR-to-UV fluxes have been corrected for Galactic 
absorption using the Schlegel et al. (1998) maps ($E_{B-V} = 0.105$) and 
the Cardelli et al. (1989) extinction curve. The conversion from 
optical/NIR magnitudes follows Fukugita, Shimasaku, \& Ichikawa (1995), 
and Bersanelli, Bouchet, \& Falomo (1991). Overplotted on the data are 
model curves of synchrotron radiation and IC scattering in 
May 2005 (solid) and ''historical" (dotted) state.  The dashed curve is 
the black body component associated with an accretion disk underlying the 
blazar}
\end{figure}

\section{Discussion}
\label{}

We have observed \3c454\ during the decaying phase of an outburst and have 
detected it with all \integral\ instruments, except PICsIT. The 3-10 keV 
flux, originating from IC scattering, is an order of 
magnitude larger than previously measured in this band (see Worrall \& 
Wilkes 1990; Comastri et al. 1997; Tavecchio et al. 2002). Both the soft 
and hard X-ray emission is brighter than observed by \integral\ 
for any other blazar, with the exception of 3C~273 (Courvoisier et al. 
2003b). The spectrum in the 3-200 keV range is well described by a single 
power-law of photon index $\sim$1.8, indicating that the spectral power 
output (in $\nu f_\nu$ representation) rises in this band, and the peak of 
the IC  component is located at the high energy limit of, or 
beyond the IBIS band, as seen in other FSRQs (e.g. S5~0836+710, 
Tavecchio et al. 2000; Pian et al. 2005a). The fluxes and spectrum 
measured by \integral\ compare well with those measured nearly 
simultaneously in the same bands by {\it Swift} (Giommi et al. 2006).  

We have reported in Fig.~3 the \integral\ data together with radio 
observations taken in April 2005, nearly one month before our campaign and 
simultaneous (within few days) mm and optical/NIR observations.  The 
optical/NIR data refer to 3 epochs corresponding to different emission 
states: we report the minimum flux recorded in optical by the OMC (May 
16.18), the intermediate flux measured by REM during the \integral\ 
observation (May 16.9), and the high flux measured by NOT on May 19.7, 
i.e. $\sim$1.2 days after the end of the \integral\ monitoring, which is 
simultaneous with a rapid X-ray flare detected by the {\it RXTE} ASM (see 
Fig.~1). We also show previous non-simultaneous multiwavelength data. 
Since \3c454\ is a bright, frequently monitored {\it CGRO} source, we 
reported its MeV-GeV spectrum observed by COMPTEL and EGRET between 1991 
and 1994 (Zhang et al. 2005; Hartman et al. 1999). The comparison between the 
\integral\ spectrum of May 2005 and the historical keV-to-MeV data 
suggests that the variability at the softer energies may be more 
pronounced than at the harder energies on a time scale of years.  Since 
the optical flux exhibits a large variability amplitude on the same time 
scale, this may indicate that the optical and soft X-ray variations are 
better correlated than those at optical and hard X-ray frequencies.  In 
our \integral\ observation we have noted this trend also on shorter time 
scales (Fig.~2). The historical SED has been fitted with synchrotron 
radiation in a homogeneous region, and IC scattering off 
synchrotron photons and external (broad line region) photons (Ghisellini, 
Celotti, \& Costamante 2002). The parameters of this fit are similar to 
the ones of Tavecchio et al. (2002).  The ``external Compton'' component 
dominates over the synchrotron self-Compton radiation, and keV-to-GeV rays 
are produced efficiently.  Using the same model, the May 2005 SED is reproduced by a brighter 
synchrotron source at mm-to-UV frequencies, and by self-Compton
scattering outside of the broad line region, at higher energies. This component
exhibits two maxima, corresponding to the first 
(\integral\ band)  and second 
order (predicted at $\sim$1 Gev) of IC scattering (Fig.~3). For the May 2005 
state we assume a size of the emitting region of $5 \times 10^{16}$ cm, 
injected luminosity of $2.2 \times 10^{44}$ erg~s$^{-1}$, magnetic field 
of 3.4 G, bulk Lorentz factor of 15, jet viewing angle of $4^{\circ}$, and 
a double power-law distribution of the relativistic electrons with a 
$\gamma_b = 700$, $\gamma_{max} = 7 \times 10^3$, and indices -2 and -4.4 
below and above $\gamma_b$, respectively. More details on this model 
solution, that represents an optimal guess of physical parameters based on the observations, 
rather than a best fit to the data, will be presented by Ghisellini (in prep.).  

The \integral\ detection of \3c454\ is in line with the finding that all 
blazars detected so far by \integral\ are also EGRET sources. \integral\ 
and {\it AGILE} observations of  EGRET blazars in outburst will allow 
us to locate accurately the peak energy of the IC component 
and to study in detail its variability and the interplay between the 
emission components.

\begin{acknowledgements}

We thank the staff at the \integral\ Mission Operation Center, and 
particularly P.~Kretschmar, for providing assistance in scheduling 
promptly these observations during non-standard satellite conditions; the 
staff at the \integral\ Science Data Center, particularly D. Eckert and N. 
Mowlavi, for data acquisition; A.D. Garau for advice in the OMC data 
analysis, and the referee, M. Gierlinski, for useful comments. 
We acknowledge the EU funding under contract HPRCN-CT-2002-00321 
(ENIGMA network). LF acknowledges the Italian Space Agency for financial 
support. RH acknowledges the ESA PECS project 98023.

\end{acknowledgements}

\end{document}